\documentclass[twocolumn,preprintnumbers,amsmath,amssymb,prb]{revtex4}

\usepackage{graphicx}% Include figure files
\usepackage{amssymb,amsmath}
\usepackage{dcolumn}% Align table columns on decimal point
\usepackage{bm}% bold math
\usepackage{color}
\usepackage{enumerate}
\usepackage{epstopdf}

\begin{document}

\title{Optically induced persistent current in carbon nanotubes}
\author{O. V. Kibis}\email{Oleg.Kibis(c)nstu.ru}
\author{M. V. Boev}
\author{V. M. Kovalev}

\affiliation{Department of Applied and Theoretical Physics,
Novosibirsk State Technical University, Karl Marx Avenue 20,
Novosibirsk 630073, Russia}

\begin{abstract}
We demonstrated theoretically that an off-resonant circularly polarized electromagnetic field can induce the persistent current in carbon nanotubes, which corresponds to electron rotation about the nanotube axis. As a consequence, the nanotubes acquire magnetic moment along the axis, which depends on their crystal structure and can be detected in state-of-the-art measurements. This effect and related phenomena are analyzed within the developed Floquet theory describing electronic properties of the nanotubes irradiated by the field.
\end{abstract}

\maketitle

\section{Introduction}
The optical control of electronic properties of solids by an off-resonant electromagnetic field based on the Floquet theory of periodically driven quantum systems (the Floquet engineering) became an emerging research area during last years~\cite{Basov_2017,Oka_2019}. Since the field frequency is far from characteristic resonant frequencies of the electron system (the off-resonant field), it cannot be absorbed by electrons in the solid and only ``dresses'' them (the dressing field), modifying their physical properties. Such a dressing of electron systems by the off-resonant field results in many fundamental effect in various nanostructures, including  semiconductor quantum wells~\cite{Lindner_2011,Kibis_2012,Pervishko_2015,Kibis_2020}, quantum rings~\cite{Kibis_2011,Kibis_2013,Sigurdsson_2014,Kibis_2015,Kozin_2018_1}, topological insulators~\cite{Rechtsman_2013,Wang_2013,Torres_2014,Calvo_2015,Mikami_2016}, graphene and related two-dimensional materials~\cite{Oka_2009,Kibis_2010,Iurov_2017,Iurov_2013,Syzranov_2013,Usaj_2014,Perez_2014,Glazov_2014,Sentef_2015,Sie_2015,Kibis_2017,Iurov_2019,Iurov_2020,Cavalleri_2020}, etc. In the present research, we developed the theory describing the interaction between electrons in such actively studied nanostructures as carbon nanotubes (CNTs)~\cite{Dresselhaus_book} and a circularly polarized electromagnetic wave (the off-resonant dressing field) propagating along the CNT axis. As a main result, it is found that the field induces the ring electric current associated with the ground electron state. Since the current-carrying state is ground, the current flows without dissipation (persistent current). The article is dedicated to the theory of this effect and organized as follows. In the second section, the model of electron-field interaction in CNTs is developed for the quantized circularly polarized field. In the third section, this model is detailed to describe the electron states in irradiated armchair CNTs. In the fourth section, analysis of the field-modified electron states and the discussion of the light-induced persistent current in the CNTs are performed. The last two sections contain the conclusion and the acknowledgements.

\section{Model}
\begin{figure}[!h]
\centering\includegraphics[width=1.\columnwidth]{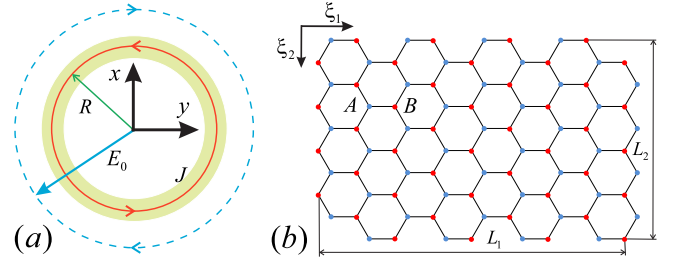}
\caption{Sketch of the system under consideration: (a) The cross-section of the CNT (the yellow ring strip) with the radius $R$ irradiated by a circularly polarized electromagnetic wave with the electric field amplitude $E_0$, which propagates along the CNT axis corresponding to the $z$ coordinate of the Cartesian coordinate system $x,y,z$ associated with the CNT. The ring red line marks the irradiation-induced persistent current $J$; (b) The graphene sheet as a basis of the $(4,4)$ armchair CNT with the radius $R=L_1/2\pi$ and the length $L_2$. The blue and red circles mark the carbon atoms from the $A$ and $B$ graphene sublattices, respectively. The axes $\xi_{1,2}$ mark the coordinate system associated with the graphene sheet. The CNT is fabricated by the rolling-up of the sheet along the $\xi_1$ axis.}
\end{figure}
We will be to consider a single-wall CNT irradiated by a plane off-resonant electromagnetic wave with the clockwise circular polarization, the electric field amplitude $E_0$ and the frequency $\omega_0$ (the dressing field), which propagates along the CNT axis corresponding to the Cartesian coordinate $z$ (see Fig.~1a). In the most general form, the Hamiltonian of an electron in the irradiated CNT reads $\hat{\cal
H}_e=[\hat{\mathbf{p}}-e\mathbf{A}/c]^2/2m_e+U(\mathbf{r})$, where
$\hat{\mathbf{p}}$ is the electron momentum
operator, $e$ is the electron charge, $m_e$ is the mass of free electron, $U(\mathbf{r})$ is the crystal periodical potential of the CNT, $\mathbf{r}$ is the electron radius vector, and
\begin{align}\label{A}
&\mathbf{A}(z,t)=(A_x,A_y,A_z)=\\
&\Big((cE_0/\omega_0)\cos\omega_0(t-z/c),\,
(cE_0/\omega_0)\sin\omega_0(t-z/c),\,0\Big)\nonumber
\end{align}
is the vector potential of the wave. Assuming that the wave length $2\pi c/\omega_0$ much exceeds the CNT length $L_2$, it is convenient to apply the unitary transformation
$\hat{\cal U}=\exp({ie}\mathbf{A}\mathbf{r}/{c\hbar})$ in order to simplify the following analysis. Then the transformed Hamiltonian $\hat{\cal H}_e$ reads
\begin{equation}\label{He}
\hat{\cal
H}_e^\prime=\hat{\cal U}^\dagger\hat{\cal H}_e\hat{\cal U} -
i\hbar\hat{\cal U}^\dagger\partial_t
\hat{\cal U}=\hat{\cal H}_0-e\mathbf{E}\mathbf{r},
\end{equation}
where the first term, $\hat{\cal H}_0=\hat{\mathbf{p}}^2/2m_e+U(\mathbf{r})$, is the electron Hamiltonian of the CNT in the absence of the field ($E_0=0$), $\mathbf{E}=-\partial_t\,\mathbf{A}/c$ is the electric field of the wave, and the last term describes the electron-field interaction within the dipole approximation.

Let us consider the CNT irradiated by the quantized field (\ref{A}), applying the approach~\cite{Kibis_2010,Iurov_2017,Iurov_2013,Kibis_2011,Kibis_2012} developed before to describe various nanostructures interacting with a quantized electromagnetic field.
Within the conventional quantum-field theory, the classical field, $\mathbf{E}$, should be replaced in the Hamiltonian (\ref{He}) by the field operator, $\hat{\mathbf{E}}=i\sqrt{2\pi\hbar\omega_0/V}\left(\mathbf{e}_+\hat{a}-
\mathbf{e}_-\hat{a}^\dagger\right)$, where $\hat{a}$ and $\hat{a}^\dagger$ are the operators of photon annihilation and creation, respectively, written in the Schr\"{o}dinger representation (the representation of occupation numbers), $\mathbf{e}_{\pm}=(\mathbf{e}_{x}\pm i\mathbf{e}_{y})/\sqrt{2}$ are the polarization vectors, $\mathbf{e}_{x,y}$ are the unit vectors directed along the $x,y$-axes of the chosen Cartesian system (see Fig.~1a), and $V$ is the quantization volume. Then the total Hamiltonian of the interacting electron-photon system in the CNT reads
\begin{equation}\label{H}
\hat{{\cal H}}=\hbar\omega_0\hat{a}^\dagger\hat{a}+\hat{\cal H}_0-ie\sqrt{2\pi\hbar\omega_0/V}\mathbf{r}\left(\mathbf{e}_+\hat{a}-
\mathbf{e}_-\hat{a}^\dagger\right),
\end{equation}
where
the first term is the field energy operator, the second term is the electron energy operator, and the third term describes the electron-field interaction within the dipole approximation.

In the absence of the irradiation, the electron energy spectrum of the CNT, $\varepsilon_\mu^{(j)}(k)$, and the corresponding electron wave functions, $\psi_\mu^{(j)}(k)$, satisfy the conventional Schr\"odinger equation, $\hat{\cal H}_0\psi_\mu^{(j)}(k)=\varepsilon_\mu^{(j)}(k)\psi_\mu^{(j)}(k)$, where $\mu$ is the quantum number labeling different electron subbands in the CNT, $j=c(v)$ is the index marking electron states in the conduction (valence) band of the CNT, and $k$ is the electron wave vector along the CNT axis~\cite{Dresselhaus_book}. To describe the electron-field interaction, let us
introduce the joint electron-photon space,
\begin{equation}\label{psi}
|\psi_\mu^{(j)}(k),N\rangle=|\psi_\mu^{(j)}(k)\rangle\otimes|N\rangle,
\end{equation}
which corresponds to an electron being in the state with the wave function
$\psi_\mu^{(j)}(k)$ and the dressing field being in the state
with the photon occupation number $N=1,2,3,...$. The basic states
of this space, $|\psi_\mu^{(j)}(k),N\rangle$, are eigenstates of the Hamiltonian of non-interacting electron-photon system and, therefore, they are orthonormal, $\langle\psi_\mu^{(j)}(k),N|\psi_{\mu^\prime}^{(j^\prime)}(k^\prime),N^\prime\rangle=\delta_{N,N^\prime}\delta_{j,j^\prime}\delta_{\mu,\mu^\prime}\delta_{k,k^\prime}$. As a consequence, one can write the total Hamiltonian (\ref{H}) as a matrix, ${\cal H}_{nm}$, where the indices $n,m$ label different basic states of the joined electron-photon space (\ref{psi}). To find the energy of the interacting electron-photon system, ${\varepsilon}$, one has to solve the secular equation with the matrix Hamiltonian,
\begin{equation}\label{det}
\mathrm{det}||{\cal H}_{nm}-{\varepsilon}I||=0,
\end{equation}
where $I$ is the unity matrix. Assuming the field frequency $\omega_0$ to be far enough from the resonant frequencies corresponding to allowed optical transitions in the CNT (the off-resonant dressing field which cannot be absorbed by electrons), the eigenenergies of the electron-photon Hamiltonian (\ref{H}) can be written as a sum,  ${\varepsilon}=N_0\hbar\omega_0+\tilde{\varepsilon}_\mu^{(j)}(k)$, where $N_0$ is the photon occupation number of the dressing field. Correspondingly, the first term of the sum is the dressing field energy, whereas the second term, $\tilde{\varepsilon}_\mu^{(j)}(k)$, is the CNT electron energy spectrum modified by the field.

It should be noted that the energy spectrum $\tilde{\varepsilon}_\mu^{(j)}(k)$ can be used to describe scattering-induced electron transitions between different eigenstates of the Hamiltonian (\ref{H}) in the conventional way if the photon energy, $\hbar\omega_0$, much exceeds the scattering-inducing broadening of the electron eigenenergies, $\hbar/\tau$, where $\tau$ is the mean free time of electrons restricted by the scattering. Otherwise, the scattering processes cannot be considered as a perturbation and must be taken into account within the Hamiltonian (\ref{H}). Thus, the field frequency must satisfy the condition $\omega_0\tau\gg1$, which is of general character for various periodically driven condensed-matter structures (see, e.g., Ref.~\onlinecite{Kibis_2017} and references therein). For CNTs fabricated with using modern nanotechnologies, this condition can be satisfied for the high field frequencies starting approximately from the upper microwave range limit.

In the following, we will assume the field to be classically intensive ($N_0\gg1$) and, therefore, one can introduce the classical field amplitude,
\begin{equation}\label{E0}
E_0=\sqrt{4\pi N_0\hbar\omega_0/V}.
\end{equation}
Then the matrix elements of the Hamiltonian (\ref{H}) in the basis (\ref{psi}) read
\begin{align}\label{Hnm}
&\langle\psi_{\mu^\prime}^{(j^\prime)}(k^\prime),N^\prime|\hat{\cal H}|\psi_\mu^{(j)}(k),N\rangle=
[N\hbar\omega_0+\varepsilon_\mu^{(j)}(k)]\times\\
&\delta_{k^\prime,k}\delta_{\mu^\prime,\mu}\delta_{N^\prime,N}
-[{D}^{(j^\prime j)}_{\mu^\prime\mu}(k) E_0]\delta_{\mu^\prime,\mu\pm1}\delta_{N^\prime,N\mp1},\nonumber
\end{align}
where
\begin{equation}\label{DD}
{D}^{(j^\prime j)}_{\mu^\prime\mu}(k)=\langle\psi_{\mu^\prime}^{(j^\prime)}(k)|i\mathbf{D}\mathbf{e}_\pm/\sqrt{2}|\psi_{\mu}^{(j)}(k)\rangle\delta_{k^\prime,k},
\end{equation}
are the matrix elements of intersubband dipole transitions, and
$\mathbf{D}=e\mathbf{r}$ is the dipole moment of electron.

It should be stressed that the Hamiltonian (\ref{H}) describes the closed system ``electrons + quantized field''. Since the energy of any closed system is conserved quantity, the Hamiltonian (\ref{H}) and their eigenstates are stationary. As a result, the stationary Schr\"odinger problem with the Hamiltonian (\ref{H}) differs mathematically from the conventional time-dependent Floquet problem for the classical periodical field (\ref{A}). However, the both description of the field --- classical and quantized --- are physically equivalent in the considered limiting case of the strong field, $N_0\gg1$. Therefore, the electron energy spectrum $\tilde{\varepsilon}_\mu^{(j)}(k)$, which can be found from the secular equation (\ref{det}) for the strong quantized field with the amplitude (\ref{E0}), exactly coincides with the Floquet (quasi)energies which can be found as a solution of the conventional time-dependent Floquet problem for the same classical field (\ref{A}). Moreover, the electron-photon eigenstates of the problem written in the basis (\ref{psi}) have the same physical meaning as the periodically time-dependent Floquet states originated from the classical field (\ref{A}). As a consequence, the stationary Schr\"odinger problem with the Hamiltonian (\ref{H}) can be treated as the Floquet problem for the quantized field (\ref{A}).  As to benefits arisen from the stationary form of the Hamiltonian (\ref{H}), the well-developed stationary perturbation theory can be applied directly to find its eigenstates and eigenenergies (see the Section IV below).

\section{Armchair carbon nanotubes in the field}
For definiteness, let us apply the model developed above to the single-wall $(n,n)$ CNT, where $(n,n)$ are the coordinates of the chiral vector defining the CNT crystal structure~\cite{Dresselhaus_book}. The $(n,n)$ CNT with the radius $R=L_1/2\pi=\sqrt{3}na/2\pi$ and the length $L_2$ can be fabricated by the rolling-up of the graphene sheet pictured in Fig.~1b along the $\xi_1$ axis, where $L_{1,2}$ are the dimensions of the graphene sheet, and $a\approx2.46$~\AA~ is the lattice constant of graphene. Since carbon atoms in the graphene sheet are positioned along the rolling-up axis $\xi_1$ in the armchair-like order, such CNTs are known also as armchair nanotubes.

Within the tight-binding approximation~\cite{Dresselhaus_book}, electronic states of graphene are described by the electron energy spectrum
\begin{align}\label{EG}
&\varepsilon^{(j)}_G(k_1,k_2)=\\\pm &t\sqrt{1+4\cos\left({\sqrt{3}k_1a}/{2}\right)
\cos\left({k_2a}/{2}\right)+4\cos^2\left({k_2a}/{2}\right)},\nonumber
\end{align}
and the electron wave function
\begin{equation}\label{FG}
\psi^{(j)}_G(k_1,k_2)=\frac{1}{\sqrt{2}}\left[\pm{\psi_A(\mathbf{k})}+{\psi_B(\mathbf{k})}\frac{f^\ast(\mathbf{k})}{{|f(\mathbf{k})|}}\right]
\end{equation}
where $f(\mathbf{k})={e^{ik_1a/\sqrt{3}}+2e^{-ik_1a/2\sqrt{3}}\cos(k_2a/2)}$,
\begin{equation}\label{Bloch}
\psi_{A,B}(\mathbf{k})=\sqrt{\frac{2}{N_a}}\sum_{\mathbf{R}_{A,B}}e^{i\mathbf{k}\mathbf{R}_{A,B}}\phi_{A,B}(\mathbf{r}-\mathbf{R}_{A,B})
\end{equation}
are the Bloch functions for the $A,B$ sublattices of graphene marked by the blue (red) circles in Fig.~1b,
$\phi_{A,B}(\mathbf{r})$ are the atomic wave functions of the carbon atoms from these two sublattices, $\mathbf{R}_{A,B}$ are the radius vectors of the atoms, $N_a$ is the total number of carbon atoms in the graphene sheet, $\mathbf{k}=(k_1,k_2)$ is the electron wave vector written in the $\xi_{1,2}$ axes marked in Fig.~1b, $t\approx3.033$~eV is the energy of interatomic electron interaction in graphene, and the two band indices $j=c,v$ correspond to the signs ``$+$'' and ``$-$'', respectively. Neglecting curvature of the CNT, the rolling-up of the graphene sheet pictured in Fig.~1b into the armchair CNT results only in the quantization of the electron wave vector along the rolling-up axis $\xi_1$. As a consequence, the quantized component of the wave vector is $k_1=2\pi\mu/\sqrt{3}an$ with $\mu=0,1,2,...2n-1$, whereas the electron wave vector along the CNT axis $\xi_2$ remains the same, $k_2=k$. As a result, Eqs.~(\ref{EG})--(\ref{FG}) yield
the CNT energy spectrum~\cite{Dresselhaus_book}
\begin{align}\label{ECNT}
&\varepsilon^{(j)}_\mu(k)=\varepsilon^{(j)}_G\left(\frac{2\pi\mu}{\sqrt{3}an},k\right)=\\
&\pm t\sqrt{1+4\cos\left({\pi\mu}/{n}\right)
\cos\left({ka}/{2}\right)+4\cos^2\left({ka}/{2}\right)}\nonumber
\end{align}
and the corresponding wave function
\begin{equation}\label{PCNT}
\psi^{(j)}_\mu(k)=\psi^{(j)}_G\left(\frac{2\pi\mu}{\sqrt{3}an},k\right).
\end{equation}
It should be noted that the CNT electron states (\ref{ECNT})--(\ref{PCNT}) with $\mu$ and $\mu\pm2n$ are physically equivalent since $\varepsilon^{(j)}_{\mu\pm2n}(k)=\varepsilon^{(j)}_\mu(k)$ and $\psi^{(j)}_{\mu\pm2n}(k)=\psi^{(j)}_\mu(k)$.
Substituting Eqs.~(\ref{ECNT})--(\ref{PCNT}) into Eq.~(\ref{DD}), the matrix elements of intersubband dipole transitions can be written within the tight-binding approach~\cite{Jiang_2004} as
\begin{align}
&{D}_{\mu^\prime\mu}^{(cv)}(k)=\mp\frac{3ne\hbar V_{AB}}{8\pi[\varepsilon_{\mu^\prime}^{(c)}(k)-\varepsilon_\mu^{(v)}(k)]}\nonumber\\
&\times\left[\frac{A_\pm(\mu,k)f^\ast(\mu,k)}{|f(\mu,k)|}+\frac{B_\pm(\mu,k)f(\mu^\prime,k)}{|f(\mu^\prime,k)|}\right]\delta_{\mu^\prime,\mu\pm1},\label{1}\\
&{D}_{\mu^\prime\mu}^{(cc)}(k)=\mp\frac{3ne\hbar V_{AB}}{8\pi[\varepsilon_{\mu^\prime}^{(c)}(k)-\varepsilon_\mu^{(c)}(k)]}\nonumber\\
&\times\left[\frac{A_\pm(\mu,k)f^\ast(\mu,k)}{|f(\mu,k)|}-\frac{B_\pm(\mu,k)f(\mu^\prime,k)}{|f(\mu^\prime,k)|}\right]\delta_{\mu^\prime,\mu\pm1},\\
&{D}_{\mu^\prime\mu}^{(vv)}(k)=\pm\frac{3ne\hbar V_{AB}}{8\pi[\varepsilon_{\mu^\prime}^{(v)}(k)-\varepsilon_\mu^{(v)}(k)]}\nonumber\\
&\times\left[\frac{A_\pm(\mu,k)f^\ast(\mu,k)}{|f(\mu,k)|}-\frac{B_\pm(\mu,k)f(\mu^\prime,k)}{|f(\mu^\prime,k)|}\right]\delta_{\mu^\prime,\mu\pm1},
\label{2}
\end{align}
where
\begin{align}
&f(\mu,k)=e^{2\pi i\mu/3n}+2e^{-\pi i\mu/3n}\cos\left(\frac{ka}{2}\right),\nonumber\\
&A_\pm(\mu,k)=e^{2\pi i\mu/3n}\left(1-e^{\pm2\pi i/3n}\right)\nonumber\\
&+2e^{-\pi i\mu/3n}\cos\left(\frac{ka}{2}\right)
\left(1-e^{\mp\pi i/3n}\right),\nonumber\\
&B_\pm(\mu,k)=e^{-2\pi i\mu/3n}\left(1-e^{\pm2\pi i/3n}\right)\nonumber\\
&+2e^{\pi i\mu/3n}\cos\left(\frac{ka}{2}\right)
\left(e^{\pm\pi i/n}-e^{\pm2\pi i/3n}\right),\nonumber
\end{align}
and
$V_{AB}=\langle\phi_A(\mathbf{r})|\hat{v}_1|\phi_B(\mathbf{r}-\mathbf{b})\rangle$
is the matrix element of the velocity operator along the $\xi_1$ axis of graphene, $\hat{v}_1$, for the electron wave functions of the two carbon atoms from the $A,B$ sublattices, which are shifted with respect to each other by the vector $\mathbf{b}=(a/\sqrt{3},0)$ (see the two atoms marked as $A$ and $B$ in Fig.~1b). To write this matrix element explicitly, one can apply the equality following directly from the graphene Hamiltonian~\cite{Saroka_2018,Hartmann_2019}, $\langle\psi_A(\mathbf{k}_{GD})|\hat{v}_1|\psi_B(\mathbf{k}_{GD})\rangle=iv_F$, where $\mathbf{k}_{GD}=(0,\pm4\pi/3a)$ is the wave vector of the Dirac points of graphene, and $v_F=\sqrt{3}ta/2\hbar$ is the Fermi velocity of electrons in graphene. Substituting the Bloch functions (\ref{Bloch}) into this equality, it yields
\begin{equation}\label{VAB}
V_{AB}=\langle\phi_A(\mathbf{r})|\hat{v}_1|\phi_B(\mathbf{r}-\mathbf{b})\rangle=i2v_F/3.
\end{equation}

\begin{figure}[h!]
\centering\includegraphics[width=1.\columnwidth]{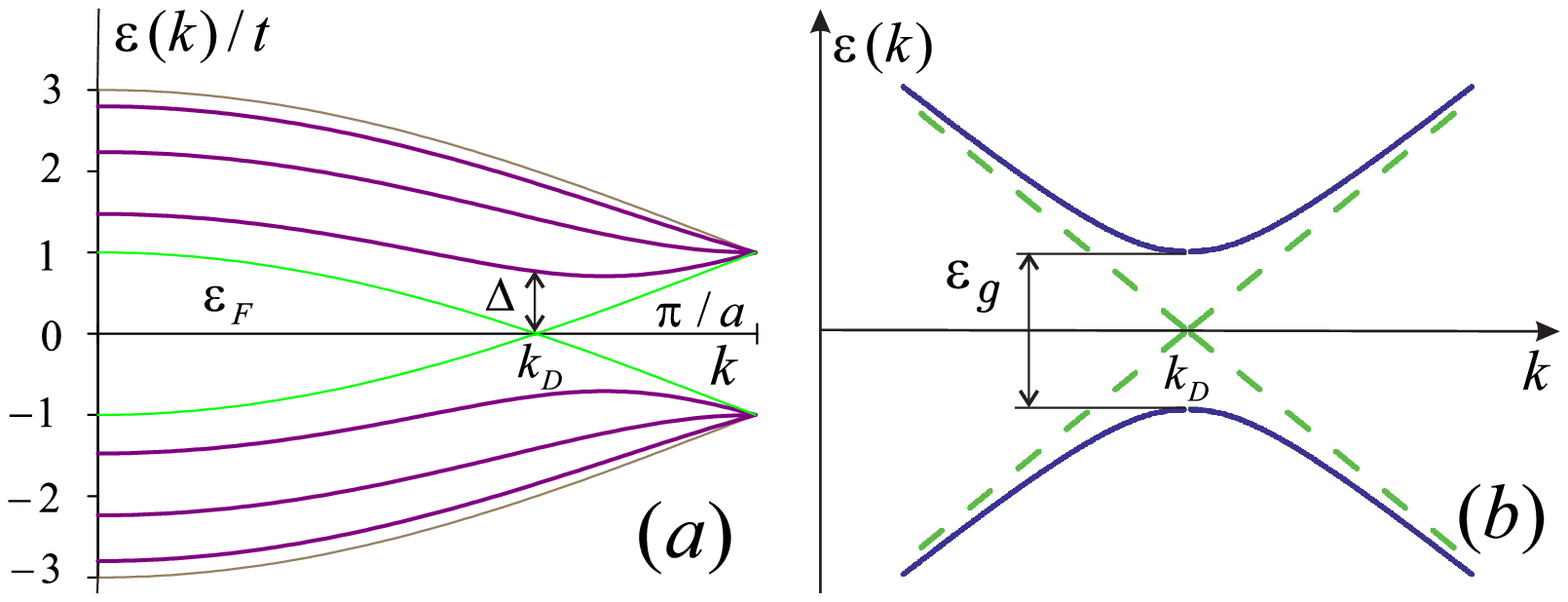}
\caption{(a) The electron energy spectrum of the $(4,4)$ armchair CNT. The Fermi energy of the electron system is $\varepsilon_F=0$, the threshold of interband optical absorption is $\Delta$ and the Dirac point wave vector $k_D=2\pi/3a$ corresponds to the band edge; (b) The electron energy spectrum of the CNT near the band edge in the absence of the dressing field (the dashed lines) and in the presence of the dressing field (the solid lines), where $\varepsilon_g$ is the field-induced band gap.}
\end{figure}
Using Eqs.~(\ref{FG}), (\ref{Bloch}), (\ref{PCNT}) and (\ref{VAB}), one can calculate the averaged angular momentum corresponding to the electron states (\ref{PCNT}) with the various indices $j$ and $\mu$,
\begin{align}\label{Lz}
&L_\mu^{(j)}=\langle{\psi}_\mu^{(j)}(k)|\hat{L}|{\psi}_\mu^{(j)}(k)\rangle=\\
&m_eR\left\langle{\psi}_{G}^{(j)}\left(\frac{2\pi\mu}{\sqrt{3}an},k\right)\Big|e\hat{v}_1\Big|{\psi}_{G}^{(j)}\left(\frac{2\pi\mu}{\sqrt{3}an},k\right)\right\rangle=\nonumber\\
&\mp \frac{2m_eRv_F\sin(\pi\mu/n)\cos(ka/2)}{\sqrt{1+4\cos(\pi\mu/{n})\cos(ka/2)+4\cos^2(ka/2)}},
\end{align}
where $\hat{L}=m_eR\hat{v}_1$ is the operator of electron angular momentum along the CNT axis and the band indices $j=c,v$ correspond to the upper and down signs ``$\mp$'', respectively. In turn, the averaged ring electric current associated with the angular momentum (\ref{Lz}) can be written as
\begin{align}\label{jj}
&J_{\mu}^{(j)}(k)=\langle{\psi}_{\mu}^{(j)}(k)|\hat{J}|{\psi}_{\mu}^{(j)}(k)\rangle=\\
&\frac{\left\langle{\psi}_{G}^{(j)}\left(\frac{2\pi\mu}{\sqrt{3}an},k\right)\Big|e\hat{v}_1\Big|{\psi}_{G}^{(j)}\left(\frac{2\pi\mu}{\sqrt{3}an},k\right)\right\rangle}{2\pi R}=\nonumber\\
&\mp\frac{ev_F}{\pi R}\frac{\sin(\pi\mu/n)\cos(ka/2)}{\sqrt{1+4\cos(\pi\mu/{n})\cos(ka/2)+4\cos^2(ka/2)}},\nonumber
\end{align}
where $\hat{J}=e\hat{v}_1/2\pi R$ is the operator of the current.  Correspondingly, the magnetic moment along the CNT axis arisen from the current (\ref{jj}) reads $M_\mu^{(j)}=J_{\mu}^{(j)}(k)\pi R^2$.

It follows from Eq.~(\ref{ECNT}) that the CNT electron energy spectrum consists of the two nondegenerate subbands with $\mu=0,n$ corresponding to the zero angular momentum (\ref{Lz}) and the set of double degenerate subbands with $\mu=n\pm\nu$ and $\nu=1,2,...,n-1$, where the signs ``$\pm$'' correspond to the electron subbands with mutually opposite orientations of the angular momentum (\ref{Lz}). Thus, the double degeneracy of the subbands should be treated as a particular case of the Kramers degeneracy. The spectrum (\ref{ECNT}) is plotted for the $(4,4)$ armchair CNT in Fig.~2a, where the bold violet lines mark the doubly degenerate subbands, and the thin lines (green and brown) mark the nondegenerate ones. The thin green lines correspond to the valence and conduction band edge subbands ($\mu=n$) which touch each other in the Dirac point with the wave vector $k_D=\pm2\pi/3a$. Thus, the electron energy structure of the armchair CNT is of metal type with the linear electron dispersion near the Dirac point. It follows from Eqs.~(\ref{1})--(\ref{VAB}) that the dipole matrix elements are zero for the optical transitions between the valence and conduction subbands with $\mu=n$. This means that the selection rules forbid  the optical transitions between the two edge subbands with the same angular momentum (\ref{Lz}) and, therefore, the threshold of optical absorption by valence electrons marked in Fig.~1a is
\begin{equation}\label{Delta}
\Delta=\varepsilon_{n+1}^{(c)}(k_D)-\varepsilon_{n}^{(v)}(k_D)=2t\sin(\pi/2n).
\end{equation}
Thus, the field (\ref{A}) cannot be absorbed under the condition
\begin{equation}\label{cond}
\hbar\omega_0<\Delta.
\end{equation}
In the following, we will assume the condition (\ref{cond}) to be satisfied, what allows to consider the field (\ref{A}) as an off-resonant dressing field. In order to apply the dipole approximation to the electron interaction with the field (\ref{A}) and consider the electron wave vector along the CNT axis, $k$, as a continuous quantity, the CNT length, $L_2$, will be assumed to satisfy the inequality
\begin{equation}\label{ineq}
a\ll L_2\ll 2\pi c/\omega_0.
\end{equation}
Substituting the CNT energy spectrum (\ref{ECNT}) and the dipole  matrix elements (\ref{1})--(\ref{VAB}) into the Hamiltonian (\ref{Hnm}) and solving the secular equation (\ref{det}) under the conditions (\ref{cond})--(\ref{ineq}), one can find the sought electron energy spectrum of the CNT, $\tilde{\varepsilon}_\mu^{(j)}(k)$, modified by the circularly polarized electromagnetic field (\ref{A}).

\section{Results and discussion}
The electron energy spectrum (\ref{ECNT}) describing the $(n,n)$ CNT with $n=4$ in the absence of the field is plotted in Fig.~2a. It follows from the plots that the conduction band edge of the CNT and its valence band edge correspond to the two electron states which are degenerate with the same zero energy at the Dirac point, $k=k_D$. Certainly, the spectrum (\ref{ECNT}) neglects the CNT curvature effects, i.e. the CNT radius $R=\sqrt{3}na/2\pi$ is assumed to exceed much the lattice constant $a$ and, correspondingly, $n\gg1$. However, the structure of electron energy spectrum (\ref{ECNT}) near the band edge, which is under consideration in the following, is the same for all armchair CNTs. This allows to use the spectrum pictured in Fig.~2a as an appropriate illustration of the band edge for any index $n$.

The band edge energies modified by the dressing field (\ref{A}) can be found from the secular equation (\ref{det}) within the conventional perturbation theory, considering the off-diagonal dipole matrix elements of the Hamiltonian (\ref{Hnm}) as a perturbation. Then the electron energy spectrum of the irradiated CNT near the band edge has the gapped structure pictured in Fig.~2b with the band gap $\varepsilon_g=\tilde{\varepsilon}_n^{(c)}(k_D)-\tilde{\varepsilon}_n^{(v)}(k_D)$, where
\begin{align}\label{Pedg}
&\tilde{\varepsilon}_n^{(j)}(k_D)={\varepsilon}_n^{(j)}(k_D)\\
&+
\sum_{\substack{{j^\prime=c,v}\\\mu^\prime=n\pm1}}\frac{E_0^2\left|{D}_{\mu^\prime n}^{(j^\prime j)}(k_D)\right|^2}{{\varepsilon}_n^{(j)}(k_D)-{\varepsilon}_{\mu^\prime}^{(j^\prime)}(k_D)+(\mu^\prime-n)\hbar\omega_0}\nonumber
\end{align}
are the band edge energies written within the second order of the perturbation theory (the quadratic approximation in the field amplitude $E_0$). Within the same perturbation theory, the electron-photon eigenstates corresponding to the band edges (\ref{Pedg}) can be written as
\begin{align}\label{PSIedg}
&|\tilde{\psi}_n^{(j)}(k_D)\rangle=|\psi_n^{(j)}(k_D),N_0\rangle+\\
&\sum_{\substack{{j^\prime=c,v}\\\mu^\prime=n\pm1}}\frac{E_0{D}_{\mu^\prime n}^{(j^\prime j)}(k_D)|\psi_{\mu^\prime}^{(j^\prime)}(k_D),N_0+n-\mu^\prime\rangle}{{\varepsilon}_n^{(j)}(k_D)-{\varepsilon}_{\mu^\prime}^{(j^\prime)}(k_D)+(\mu^\prime-n)\hbar\omega_0}.\nonumber
\end{align}
Substituting the unperturbed energy spectrum (\ref{ECNT}) and the dipole matrix elements (\ref{1})--(\ref{2}) into the band edge energies (\ref{Pedg}), the optically induced gap reads
\begin{equation}\label{Eg}
\varepsilon_g=\left(\frac{eRE_0}{2}\right)^2\left[\frac{1}{\Delta-\hbar\omega_0}
+\frac{1}{\Delta+\hbar\omega_0}\right]\sin\left(\frac{\pi}{2n}\right).
\end{equation}
Thus, the circularly polarized irradiation turns the metallic CNT into insulator with band gap (\ref{Eg}). It should be noted that the similar metal-insulator transition induced by a circularly polarized field takes place also in graphene~\cite{Kibis_2010}. However, the field-induced band gap in graphene is linear in the electric field amplitude, $E_0$, whereas the gap (\ref{Eg}) in CNTs depends on the amplitude quadratically. It should be noted also that Eq.~(\ref{Eg}) for $\omega_0=0$ describes the bandgap induced by a stationary electric field $E_0$ directed perpendicularly to the CNT axis.
\begin{figure}[!ht]
\includegraphics[width=.8\columnwidth]{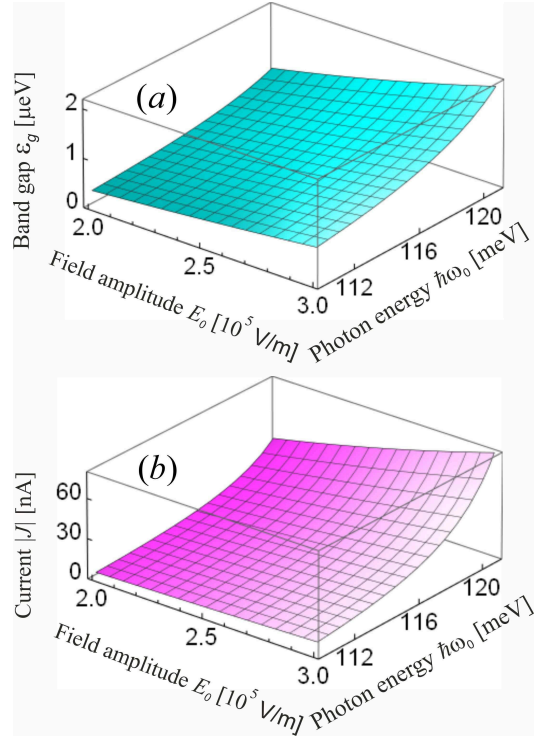}
\caption{The dependencies of the optically induced band gap $\varepsilon_g$ and persistent current $J$ on the field amplitude $E_0$ and the photon energy $\hbar\omega_0$ for the $(n,n)$ armchair CNT with $n=75$ (the CNT radius is $R\approx5.1$~nm).}
\end{figure}

Electron motion in a CNT can be considered as a superposition of translational motion of an electron along the CNT axis with the wave vector $k$ and its rotation about this axis with the angular momentum (\ref{Lz}). Using Eqs.~(\ref{Pedg})--(\ref{PSIedg}) and Eqs.~(\ref{EG})--(\ref{VAB}), we arrive at the ring electrical current, $J$, corresponding to this rotation and associated with the band edge states (\ref{Pedg})--(\ref{PSIedg}),
\begin{align}\label{j}
&J=
\langle\tilde{\psi}_n^{(j)}(k_D)|\hat{J}|\tilde{\psi}_n^{(j)}(k_D)\rangle=J_{n}^{(j)}(k_D)+\\
&\sum_{\substack{{j^\prime=c,v}\\\mu^\prime=n\pm1}}
\frac{E_0^2\left|{D}_{\mu^\prime n}^{(j^\prime j)}(k_D)\right|^2J_{\mu^\prime}^{(j^\prime)}(k_D)}{[{\varepsilon}_{n}^{(j)}(k_D)-{\varepsilon}_{\mu^\prime}^{(j^\prime)}(k_D)+(\mu^\prime-n)\hbar\omega_0]^2}.\nonumber
\end{align}
In the absence of irradiation, the ring current (\ref{jj}) corresponding to the band edge is zero, $J_{n}^{(j)}(k_D)=0$.
Therefore, the irradiation-induced contribution of the currents arisen from the states $\mu=n\pm1$, which is described by the second line of Eq.~(\ref{j}), is crucial for the discussed effect.
It should be noted that both the angular momenta (\ref{Lz}) and the currents (\ref{jj}) corresponding to the two states $\mu=n\pm1$ are equal and opposite directed, $L_{n+1}^{(j)}(k)=-L_{n-1}^{(j)}(k)$ and $J_{n+1}^{(j)}(k)=-J_{n-1}^{(j)}(k)$. However,  the overall contribution of these states into the edge state current (\ref{j}) is nonzero since
the interaction of the circularly polarized field (\ref{A}) with them is different, ${D}_{n+1,n}^{(j^\prime j)}(k_D)\neq{D}_{n-1,n}^{(j^\prime j)}(k_D)$. As a consequence, the total current (\ref{j}) differs from zero and reads
\begin{eqnarray}\label{j0}
J&=&\left(\frac{eRE_0}{2}\right)^2\frac{ev_F}{2\pi R}\cos\left(\frac{\pi}{2n}\right)\nonumber\\
&\times&\left[\frac{1}{(\Delta-\hbar\omega_0)^2}
-\frac{1}{(\Delta+\hbar\omega_0)^2}\right].
\end{eqnarray}
Since the current (\ref{j0}) does not depend on the band index $j$, it is the same for both the conduction and valence band edges. It should be noted also that the ring current (\ref{j0}) was assumed to be produced by the field (\ref{A}) with the clockwise circular polarization. In the case of counterclockwise polarization, the current (\ref{j0}) pictured schematically in Fig.~1a changes its direction to the opposite.

In an intrinsic (undoped) armchair CNT, electrons fill the valence band, whereas the conduction band is empty. Certainly, the filled valence band cannot produce any current. Let us place an extra electron into the conduction band edge state, $\tilde{\varepsilon}_n^{(c)}(k_D)$, which is ground for conduction electrons.  Since the CNT valence band is filled by other electrons, the Pauli principle forbids transitions of the electron to lower energy states there. As a consequence, the ring current (\ref{j0}) produced by the electron is disspationless (persistent) since there is no way to dissipate it with decreasing electron energy. It should be noted also that the dissipationless nature of the current (\ref{j0}) follows directly from the condition (\ref{cond}) which forbids the field absorption by electrons. Under this condition, there is no energy transfer from the field to electrons. As a consequence, the energy conservation law forbids the Joule heating associated with any current induced by the field (\ref{A}) under the condition (\ref{cond}). If the CNT is filled by $n_e$ conduction electrons (arisen, e.g., from the gate voltage or doping), the total persistent current produced by them for the zero temperature and the Fermi energy $\varepsilon_F\ll\Delta$ is just the current (\ref{j0}) multiplied by $n_e$.
As a consequence, the irradiation results in the CNT magnetization with the magnetic moment directed along the CNT axis,
\begin{equation}\label{Mz}
M_z=n_eJ\pi R^2.
\end{equation}
It follows from the axial symmetry of CNT that the magnetic moment along the CNT axis (\ref{Mz}) is just product of the considered persistent current and the cross-section area of CNT. It should be noted also that the Coulomb interaction can modify the persistent current. Therefore, the application of Eq. (\ref{Mz}) to the many electron case should be considered as a first approximation.

It follows from Eqs. (\ref{Eg}) and (\ref{j0}) that the field-induced band gap $\varepsilon_g$ and the persistent current $J$ increase with increasing the CNT radius $R=\sqrt{3}na/2\pi$. Therefore, the large-radius CNTs should be used in experiments. Since the modern nanotechnology allows to fabricate single-wall CNTs with the maximal radius around 5 nm (see, e.g., Ref.~\onlinecite{Ma_2009}), the dependence of the band gap (\ref{Eg}) and the current (\ref{j0}) on the irradiation is plotted in Fig.~3 for the $(n,n)$ CNT with $n=75$. It follows from the plots that the infrared irradiation of the intensity $\sim$~kW/cm$^2$ induces the gap $\varepsilon_g\sim\mu$eV and the current $J$ of sub-microampere scale. Although such a gap looks too small for easy detecting, the single-electron current plotted in Fig.~3b is macroscopically strong due to the giant Fermi velocity of graphene, $v_F\approx10^{8}$~cm/s, taken into account by Eq.~(\ref{j0}). Therefore, the corresponding magnetization (\ref{Mz}) is also strong and can be detected in the conventional magnetic measurements based, e.g., on the superconducting quantum interference devices (SQIDs) or torque magnetometry~\cite{Matthews_2004}. It should be noted  that the electron-phonon interaction can destroy the discussed persistent current. Therefore, the measurements should be performed at low temperatures, when the electron-photon interaction is suppressed.

It follows from the present analysis that the current (\ref{j0}) has the same physical nature as a persistent current in quantum rings induced by a circularly polarized field~\cite{Kibis_2011,Kibis_2013}. Indeed, the both currents originate from the different interaction between a circularly polarized field and electron states with mutually opposite orientations of angular momentum along the field axis. Such an asymmetry of electron-field interaction results in the optical analog of the Aharonov-Bohm effect in various ring-shaped mesoscopic systems~\cite{Sigurdsson_2014,Kibis_2015}, which manifests itself, particularly, in the discussed persistent currents. It should be noted that the previous studies of ring-shaped structures cited above were based on the simplest model of parabolic dispersion of charge carriers. In contrast to them, the  persistent current (\ref{j0}) and related phenomena depend strongly on the complicated band structure of CNTs, which is taken accurately into account within the developed theory. It should be stressed also that the persistent current (\ref{j0}) differs crucially from the usual photovoltaic currents. Indeed, any photovoltaic effect appears due to the light absorption by electrons, whereas the considered field (\ref{A}) cannot be absorbed under the condition (\ref{cond}). In the same reason, the optically induced magnetization $M_z$ associated with the ring current (\ref{j0}) differs from the conventional inverse Faraday effect which also needs the light absorption to transfer angular momentum from the field to electrons.

In contrast to the conventional Aharonov-Bohm effect~\cite{Aharonov_1959,Chambers_1960} and the related persistent currents induced by a magnetic flux in various ring-shaped structures~\cite{Webb_1985,Timp_1987,Vegvar_1989,Wees_1989}, the magnetic flux produced by the field (\ref{A}) through the CNT cross-section is zero. Therefore, the microscopical mechanisms of the persistent current (\ref{j0}) and the persistent current induced by a magnetic field differ from each other. However, there is a close analogy between these phenomena. Physically, the analogy is based on the broken time-reversal symmetry of the system, which takes place due to a circularly polarized electromagnetic field as well as a stationary magnetic one. Indeed, a circularly polarized field is non-invariant with respect to the time reversal since it turns clockwise polarized photons into counterclockwise polarized ones and vice versa. This is why a circularly polarized field acts in CNTs similarly to a stationary magnetic one in the broad range of various phenomena. In particular, the optically induced band gap in the Dirac point (\ref{Eg}) can be opened by a magnetic field directed along the CNT axis as well~\cite{Portnoi_2008,Portnoi_2009}. Next, it follows from analysis of the secular equation (\ref{det}) that the circularly polarized field (\ref{A}) results also in the splitting of the doubly generate subbands (marked by the bold blue lines in Fig.~2a), which is quadratic in the field amplitude $E_0$. Since the ring currents (\ref{jj}) corresponding to these subbands are mutually opposite, the subbands are degenerate with respect to the mutually opposite orientations of electron angular momentum along the CNT axis. Therefore, such an optically induced splitting of them should be treated as an optical analog of the Zeeman effect.

Since the irradiation-induced effects discussed above follow directly from the broken time-reversal symmetry of the electron-field system, they take place in CNTs with any crystal structure as well. However, CNTs with the crystal structure devoid of an inversion center (the chiral CNTs)~\cite{Dresselhaus_book} should be noted specially. It follows from the well-known Kramers theorem that the symmetric dependence of the electron energy on the electron wave vector in solids, $\varepsilon(\mathbf{k})=\varepsilon(-\mathbf{k})$, is the direct consequence of one of two symmetries: The inversion symmetry of crystal structure and the time-reversal symmetry (see, e.g., Ref.~\onlinecite{Callaway_book}). If the both symmetries are broken, the asymmetrical energy spectrum of electrons, $\varepsilon(\mathbf{k})\neq\varepsilon(-\mathbf{k})$, appears. Particularly, such an asymmetrical spectrum takes place in various nanostructures without an inversion center exposed to a stationary magnetic field, including asymmetric semiconductor quantum wells in the presence of an in-plane magnetic field~\cite{Kibis_1999}, magnetic edge states in two-dimensional electron systems~\cite{Kibis_2002} and chiral CNTs subjected to a magnetic field directed along the CNT axis~\cite{Kibis_1993,Kibis_2001,Kibis_2002_1}. As a consequence of the asymmetric electron dispersion, the unusual photovoltaic effects~\cite{Tarasenko_2008,Tarasenko_2011} and electron-phonon effects~\cite{Kibis_1999,Kibis_2001,Kibis_2002} induced by a magnetic field appear. Since a circularly polarized electromagnetic field acts similarly to a stationary magnetic one, the asymmetry of the electron spectrum is expected in irradiated chiral CNTs as well. Microscopically, this follows from the fact that the conserved physical quantity in CNTs with chiral (helicoidal) crystal structure is the combination of the electron angular momentum along the CNT axis and the electron momentum, $k$, along the same axis~\cite{Kibis_2005,Maksimenko_2007}. Since the circularly polarized field (\ref{A}) splits the electron states with mutually opposite orientation of the electron angular momentum, this splitting will lead to the asymmetrical electron dispersion along the CNT axis, $\varepsilon(k)\neq\varepsilon(-k)$. As a consequence,
the mentioned above effects induced by a magnetic field in nanostructures without an inversion center are expected in the chiral CNTs irradiated by a circularly polarized field as well. However, the detailed analysis of these effect goes beyond the scope of the present article and will be done elsewhere.

\section{Conclusion}
The Floquet theory of carbon nanotubes (CNTs) driven by an off-resonant circularly polarized electromagnetic wave propagating along the CNT axis is developed. It is demonstrated that the wave acts similarly to a stationary magnetic field directed along the same axis. In particular, the wave opens the gap between the conduction and valence bands of armchair CNTs (the optical analog of metal-insulator transition induced by a magnetic field) and splits the degenerate CNT subbands corresponding to mutually opposite orientations of angular momentum along the CNT axis (the optical analog of the Zeeman effect). As a main result, it is shown that the wave induces the band edge state associated with the ring electrical current. Since the band edge state is ground for conduction electrons, the current flows without dissipation and is persistent (the optical analog of the Aharonov-Bohm effect in mesoscopic rings). It should be stressed that the single-electron persistent current is macroscopically strong due to the giant Fermi velocity in graphene-related materials. Therefore, the corresponding magnetization of the CNT is also strong and can be detected in the conventional magnetic measurements based, e.g., on the superconducting quantum interference devices (SQIDs) or torque magnetometry.

\begin{acknowledgments}
The reported study was funded by the Russian Foundation for Basic Research (project 20-02-00084). We thank the Russian Ministry of Science and Higher Education (project FSUN-2020-0004).
\end{acknowledgments}

\end{document}